# Alternative Detection Methods for Highest Energy Neutrinos

Rolf Nahnhauer [a]

[a] Deutsches Elektronen Synchrotron, Platanenallee 6, D-15738 Zeuthen, Germany

Several experimental techniques are currently under development, to measure the expected tiny fluxes of highest energy neutrinos above $10^{18}$ eV. Projects in different stages of realization are discussed here, which are based on optical and radio as well as acoustic detectors. For the detection of neutrino events in this energy range a combination of different detector concepts in one experiment seems to be most promising.

## 1. Introduction

Today the detection of high energy neutrino events from astrophysical sources is based on optical Cerenkov telescopes using water or ice as target and detection medium. The present-generation detectors are operated already for several years or are close to completion. These experiments are calibrated by measuring the atmospheric neutrino flux known today with ~20 % precision. The sensitivity of the detectors is optimized for the TeV-PeV energy region. They are now considered to be prototypes of the next-generation devices of ~1 km$^3$ volume [1].

However even these gigantic detectors will measure less than one event per year from cosmogenic (GZK) neutrinos [2,3], which are due to interactions of high energy protons with the cosmic 2.7 °K photon background radiation (see Fig.1) and might serve as a calibration source at highest energies.

The construction of larger optical detectors is a difficult and expensive task. The effective light attenuation length in water and ice is less than 100 m, which fixes the distance between optical

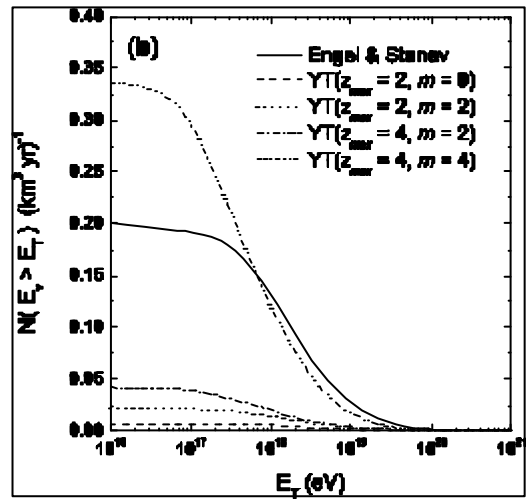

Fig. 1: Integrated GZK neutrino event rates in ice as function of the detector energy threshold, taken from [3]

modules for a given energy domain.

However, highest energy neutrinos produce also radio [4] and/or acoustic [5] signals in several materials. Under certain conditions both have attenuation lengths of about 1000 m [6]. This may allow to do measurements with much less sensors per unit volume compared to the optical approach.

## 2. Super Optical Cerenkov Telescopes

At the $10^{18}$ eV energy scale neutrino induced showers and muons produce so much light, that photo-multipliers are triggered over hundreds of meters despite the strong light attenuation. Accepting a high energy threshold of the telescope one could therefore increase the distances between the optical sensors beyond those in present projects. In such a scheme one would take advantage from the well developed technology of optical module construction, deployment and data transfer.

A corresponding proposal has been discussed in [7]. It was suggested to add a certain number of additional rings of strings of optical modules around the present IceCube detector. IceCube-Plus (1 ring at 300-500 m distance with 13-18 strings) would allow to double the number of detectable GZK neutrinos. HyperCube (4 additional rings with 81 strings and 500 m module distance) could detect ~7 GZK neutrino events per year

## 3. Radio Cerenkov Detectors

It was outlined already in 1962 that electron-photon cascades in matter should emit coherent radiation of radio waves due to a negative charge excess in the corresponding shower [4] (see Fig.2). The coherence condition demands that the emission region (characterized by the cascade diameter) is smaller than the wavelength of the emitted signal. This is fulfilled for frequencies of .1-1 GHz.

The predicted effect was confirmed only a few years ago by measurements stopping a photon beam in silica sand [8]. Several experiments started even earlier to search for radio emission of neutrino induced showers at highest energies.

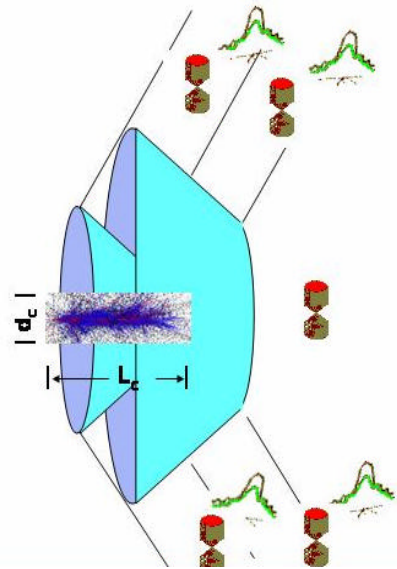

Fig. 2: Schematic view of radio Cerenkov radiation from a neutrino induced particle cascade

### - The RICE experiment

The attenuation length of radio signals of .1-1 GHz in cold Antarctic ice is larger than a km. This is used in the Radio Ice Cerenkov Experiment located at the South Pole. During 1996-98, receivers and transmitters were deployed at 100-300 m depth in holes drilled for the deployment of AMANDA optical strings. An additional dry hole allows systematic studies of ice properties with respect to radio waves. "Radioglaciology" studies concern e.g. index of refraction and absorption vs. depth. The reflection of transmitter signals from the bottom rock was observed after 5.6 km signal propagation through the polar ice.

The presently best limits for neutrino fluxes between $10^{17}$-$10^{20}$ eV were published from only 3500 hours of observation time [9] (see Fig. 3). An extension of the experiment to a 10-20 times larger volume is in the proposal stage.

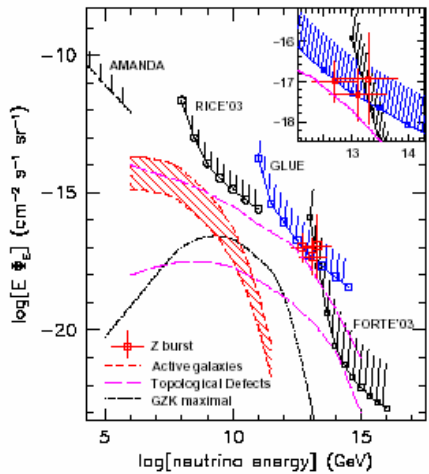

Fig. 3: Neutrino flux limits for the radio detection experiments RICE, GLUE and FORTE (from [12]).

### - The SALSA project

Salt is another material which is transparent to radio waves in the several hundred MHz region with ~km attenuation length [10]. Big salt domes exist at many locations in the world including the USA and middle Europe. They allow to install neutrino detectors of the few ten cubic kilometer scale. The SALSA project describes a corresponding device, with up to 1000 radio detectors installed in an 8 km$^3$ volume. Monte Carlo calculations demonstrate the capability of the detector to observe several 10 GZK neutrino events per year [11].

### - The GLUE experiment

Following an idea of Askaryan from 1962 [4] the Goldstone Lunar Ultra high energy neutrino Experiment uses two radio antennas to look for radio signals produced by neutrino induced cascades in the regolith at the moon surface. The target volume is in this case ~10$^5$ km$^3$, the observation time however until now only 120 hours [12]. This leads to the present best limit on the neutrino flux in the 10$^{20}$-10$^{22}$ eV energy region excluding certain Z-burst models, as can be seen from Fig. 3.

### - The FORTE experiment

Another attempt to use a big target volume was made in the Fast on Orbit Radio Transient Experiment searching from a satellite for radio signals possibly emitted from Greenland ice. No events were observed among 4·10$^6$ triggers taken during 1997-99 [13]. Figure 3 reflects that this gives the presently best limit on the neutrino flux above 10$^{22}$ eV.

### - The ANITA project

The Antarctic Impulsive Transient Antenna is payload to a balloon flight around Antarctica [14]. NASA funding started in 2003 for a first launch in 2006. A first test measurement was performed with ANITA-lite in December 2003. Radio signals were collected with a prototype detector during 18 days at float altitude [15]. The initial scan of the data revealed no obvious neutrino signal. Calibration tests were performed using ground antenna transmitters. The experiment aims to be sensitive to detect several GZK neutrino interactions during the foreseen observation time.

## 4. Air Shower Radio Signals

Radio signals from air showers were reported to be measured about 40 years ago [16]. Present ideas suggest, that they arise from the deflection of electron positron pairs in the earth magnetic field.

This leads to highly beamed pulses of synchrotron radiation with coherent emission at frequencies as low as a few hundred MHz [17].

There are several advantages to use radio detectors for the observation of air showers: the full-day observation time, the possibility to study the shower development and the low cost. The efficient identification of $\nu_e$ and $\nu_\tau$ induced showers is possible by measuring the shower direction and depth of origin in the atmosphere.

### - KASCADE/GRANDE – AUGER

Measurements of this type are done in the LOPES project [18]. Ten radio antennas were added last year to the KASCADE detector which delivers trigger signals for air showers with energies above $10^{16}$ eV. After noise reduction studies a first candidate event has been reported [19]. During the coming years ~100 antennas will be installed at KASCADE-GRANDE. This will allow to search for possible radio signals from O(1000) events with energies above $10^{17}$ eV. If this measurements are successful, radio detectors are planned to be added to the AUGER detector improving its neutrino detection capability considerably [19].

### - CODALEMA

Six antennas at the Nançay Radio Observatory form the Cosmic Detector Array of Logarithmic Electro Magnetic Antennas [20]. They look for radio emission of cosmic air showers in the 10-100 MHz region. About 9000 events have been collected with more than 10% of them giving coincident signals of all six antennas. No neutrino events have been identified until now.

## 5. Acoustic Detectors

The Thermo-Acoustic Model formulated already in 1957 [5] is based on the assumption, that the energy produced by particle cascades in a small volume of matter during a short time will overheat that volume, leading to a pressure pulse. The amplitude of this pulse measures the cascade energy. The pulse propagates like a flat disk, allowing to determine the incoming cascade direction. The frequency of the pressure wave is in the 10-100 kHz range and is related to the transversal dimension of the particle shower (see also Fig. 4).

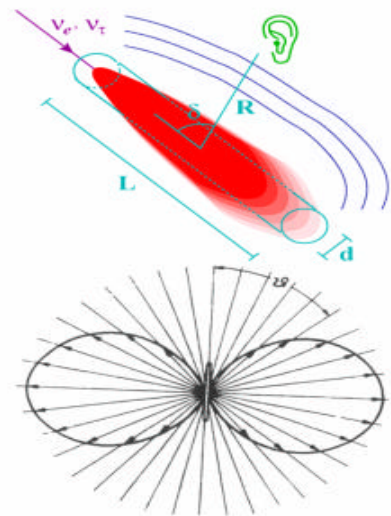

Fig.4. Illustration of the creation of a pressure pulse from an incoming neutrino induced particle cascade

At the end of the seventies several theoretical and experimental studies, done in particular for the DUMAND project, confirmed basic properties of the Thermo-Acoustic approach [21].

| group | experiment | location | activities |
|---|---|---|---|
| Stanford | SAUND | Bahamas | data taking, signal processing, calibration, simulation |
| Moscow INR1 | AGAM MP10 | Kamtschatka, Black Sea | signal processing, calibration, simulation |
| Moscow INR2, Irkutsk | Baikal | Lake Baikal | signal processing, noise studies |
| Moscow ITEP | Baikal, Antares | Lake Baikal, Mediterran. | detector R&D, accel. tests, in situ tests at Baikal, signal processing, noise studies |
| Marseille | Antares | Mediterran. | detector and installation R&D, calibration, noise studies, simulation, |
| Erlangen | Antares | Mediterran. | detector R&D, accel. tests, calibration, simulation |
| Rom, Catania | NEMO | Mediterran. | installation R&D, noise studies, simulation |
| Un. Kingdom | Rona, Antares | Scotland, Mediterran. | simulation, signal processing, calibration |
| U. Texas | Salt Dome | Hockley | detector R&D, attenuation studies, material stud. |
| Berkeley, DESY, Uppsala | IceCube | South Pole | detector R&D, accel. tests, material studies, simulation |

Table 1: Acoustic projects at different sides

Again the hope was, that due to the weak attenuation of sound waves in water, it would be possible, to survey much larger volumes than with optical sensors. However the sensors available at that time had very high energy thresholds and noise problems in the deep sea turned out to be more serious than originally expected.

During the last years acoustic particle detection in water has revived and in addition R&D projects started to show the even better capabilities of this technique for solid materials. It is impossible to describe all ongoing projects in detail (see Table 1). Instead a few special cases will be discussed in the following.

### - SAUND

A Study of Acoustic Ultrahigh-energy Neutrino Detection was done using seven hydrophones of the AUTEC military array near the Bahamas. Hydrophones are located at ~1600 m depth with distances of ~1.5 km. $651·10^6$ events have been collected in 195 days.

Careful simulations have shown the importance of refraction for track reconstruction. After all cuts for background reduction, no events remained. For the first time a neutrino flux limit was calculated using data from an acoustic detector [22]. The limit for energies above $10^{21}$eV is still not competitive with results from radio detectors but will be improved in a next stage of the experiment using many more hydrophones.

### - ITEP Moscow

Similar to what has been done 20 years ago, tests of the Thermo-Acoustic Model were performed by a group at ITEP Moscow, stopping an intense beam of 125 and 200 MeV protons in a water tank [23]. In addition to the parameters studied previously, the sound wave propagation was measured in detail. In Fig. 5 it is demonstrated that the proton beam produces a cylindrical wave front plus two spherical ones, originating from the entrance point at the container wall and the Bragg peak where the photon stops.

The group has developed own hydrophones with excellent sensitivity. They are used for measurements in and below the ice cover of Lake Baikal and are foreseen for measurements within the ANTARES acoustic R&D studies.

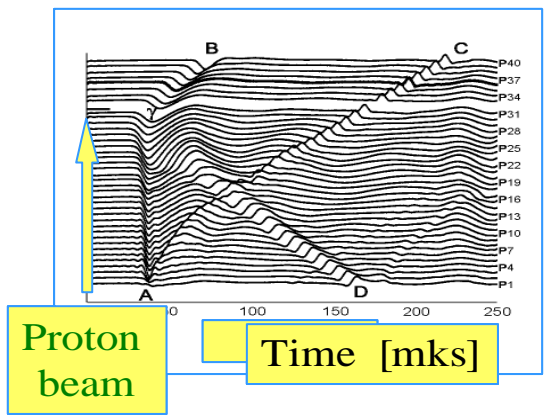

Fig. 5: Sound wave propagation in time produced by a 200 MeV proton beam stopped in water (from [23])

### - Marseille

Eight hydrophones of the TREMAIL military array located near to the ANTARES side were used to take three Gbytes of data for noise condition studies. Acoustic strings are designed and corresponding hydrophones tested. An autonomous data acquisition system is under construction [24].

### - Erlangen

The group is active in many directions of the field: simulation of acoustic signals due to intense proton beams and high energetic particle showers, sound propagation in matter and sound excitation of piezo sensors, development of cheap hydrophones as well as general tests of the Thermo-Acoustic Model [25]. During a test run at the TSL Uppsala proton synchrotron precise studies of the temperature dependence of the signal amplitudes have been performed. Preliminary results are shown in Fig. 6. Agreement with the model predictions was found.

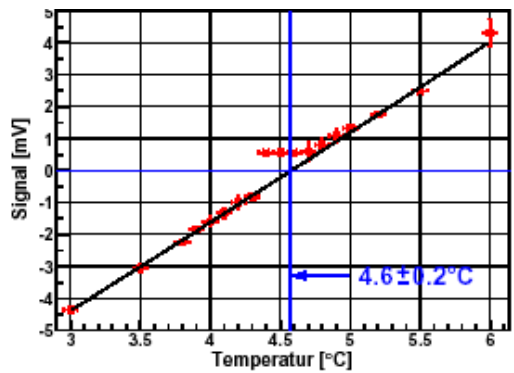

Fig. 6: Dependence of the signal amplitude on the water temperature for 177 MeV protons dumped in water (preliminary results, from [25]).

### - United Kingdom

A 1000-hydrophone array has been simulated for a large detector in the Mediterranean. The hydrophones are arranged in triplet clusters with a cluster spacing of 300 m. Good sensitivity is expected above energies of $10^{19}$ eV [26].

Using hydrophones of the Rona military array first in-situ noise studies and signal tracing attempts are under way [27].

### - Acoustic signals in salt

First measurements of the propagation of acoustic waves in natural salt have been done by a group of the University of Texas at Austin [28] in the Hockley mine. A problem was the coupling of transmitters and sensors to the salt in boreholes of several meters depth. Having solved that, longitudinal waves and shear waves have been observed as expected. Presently studies of the attenuation length in dependence on the signal frequency are ongoing.

### - *Acoustic signals in ice*

Acoustic signals in ice are expected to be much larger than in water, due to the higher velocity of sound in ice. This quantity depends however on the temperature, and therefore the location of the detector is of importance [29]. At -45 °C - a typical temperature of South Pole ice at 1 km depth - the pressure amplitude produced by a particle induced shower of a certain energy should be about ten times larger than in water at room temperature. The expected attenuation length is about a km under these conditions.

A group of institutes (University of Berkeley, DESY, Uppsala University) involved also in the AMANDA and IceCube projects have joined their activities in a feasibility study for a large acoustic experiment at the South Pole [30]. Cheap acoustic detectors have been developed at DESY (see Fig. 7) and tested at TSL Uppsala, dumping 177 MeV protons with varying intensity in an ice block at various ice temperatures. The results confirmed the model predictions. Increasing sensitivity for decreasing temperatures was measured down to -28°C. The energy threshold (10 PeV at 1 m distance) was due to internal and environmental noise and can be reduced in future. For the simulation of the detector response in ice at the South Pole it is essential to know the in-situ noise conditions. A corresponding measurement is planned for the near future, deploying 2-3 acoustic test strings with transmitters and sensors at appropriate depth.

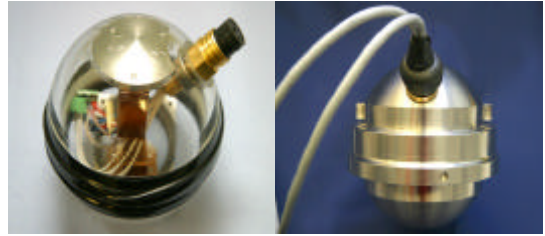

Fig. 7: Two acoustic detectors developed at DESY for the deployment in ice. Both contain two sensors and corresponding front-end electronics.

## 6. Summary and conclusions

For the observation of the tiny neutrino fluxes above $10^{18}$ eV optical detection techniques seem to reach their limits. Radio detectors start to become sensitive in this energy region. Acoustic detectors have a big revival but still need R&D efforts.

Further progress in the field would profit from world wide cooperation.

A combination of different detection methods in one experiment might be the most promising approach for the save detection of neutrino signals at highest energies.

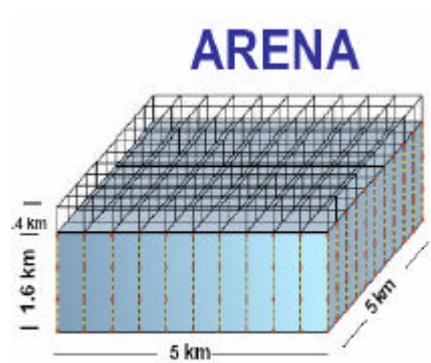

Fig 8: Sketch of a large volume radio-acoustic detector, ARENA: "Acoustic and Radio EeV Neutrino-detection Array"

Dedicated detailed proposals for large scale experiments will help to understand the real problems and challenges of the subject. An example for a combined radio-acoustic detector is given in Fig. 8. With an effective volume of more than 70 km$^3$ dependent on the target material tens of GZK neutrinos per year could be studied. The design is based on ~1000 radio receivers and ~4000 acoustic sensors located on 11x11 strings of 2 km length.

**Acknowledgement**

For helpful discussions and unpublished material I would like to thank:
G. Anton, S. Barwick, D. Besson,
S. Boser, H. Falke, M. Fink, K. Graf,
G. Gratta, A. Hallgren, F. Halzen,
A. Haungs, U. Katz, P. Kollmannsberger,
S. Kuch, J. Learned, B. Price,
A. Rostovtzev, K. Salomon, D.Saltzberg,
D. Seckel, C. Spiering, J. Stegmaier,
L. Thompson, J. Vandenbroucke